# Synthesis of Bi-based superconductor by microwave-assisted hydrothermal method

**R. G. Lima[1], V. D. Rodrigues[1], C. L. Carvalho[1], S. R. Teixeira[2], A. E. Souza[2] and R. Zadorosny[1]**

[1]Depto de Física e Química, Faculdade de Engenharia de Ilha Solteira, Univ Estadual Paulista - UNESP, Caixa Postal 31, 15385-000, Ilha Solteira, SP, Brazil

[2]Depto de Física, Química e Biologia, Faculdade de Ciências e Tecnologia, Univ Estadual Paulista - UNESP, Presidente Prudente, SP, Brazil

rafazad@yahoo.com.br

**Abstract**. In this work we studied the synthesis of BSCCO-2212 superconducting phase associating a quite similar method developed by Pechini (PM) with the microwave-assisted hydrothermal method, MAH. To study the influence of MAH on the properties of BSCCO system, we synthesized two samples by such method. For one sample we used carbonates (CMAH) and for the other one we used nitrates (NMAH) as chemical reagents. We also produced a reference sample (REF) just using carbonates by PM to compare their morphological and superconducting properties. The structural properties of the samples were analyzed by scanning electron microscopy and X-ray diffraction. It can be noted that the Bi-2212 phase is predominant in all samples; MAH samples present larger grains than the REF one. However, we can observe that the magnetic behaviour of NMAH is closer to that presented by REF.

## 1. Introduction

Studies of new routes to synthesized oxide superconductors are focused on the production of materials with higher $T_c$ and $J_c$ [1,2]. Chemical methods as sol-gel, Pechini, coprecipitation and hydrothermal have been applied on the synthesis of superconducting materials due to the homogeneity to produce samples. [1,3] Thereby, on the last years, the use of microwaves in the synthesis of a variety of materials has been raised in areas as chemistry, condensed matter physics and materials engineering. In some researches, the use of domestic microwave-oven in scientific activities has been of great interest due to its simplicity and low cost operation [4]. It is interesting to emphasize that several inorganic oxides, including the CuO, absorbs micro-wave radiations as those produced by domestic microwave-oven, i.e., with frequency of 2.45GHz. Baghurst and coworkers [6] were the pioneers on the application of microwaves to synthesize mixtures of metallic oxides with superconducting properties. In such samples, the heating process begins in the inner of the copper oxides and then the heat is transferred to the vicinity. Thus, the crystallization temperature of the material is reached faster than in conventional heating processes, saving time and energy. [5,6]

Particularly, in the microwave-assisted hydrothermal method, MAH, those waves interact with the solution and part of the electromagnetic energy is converted in thermal energy. Thus, this heat is generated in the inner solution and homogeneously propagates to the entire system promoting an extremely fast crystallization kinetics. [7,8] In contrast with the conventional heating process, there are several advantages in use MAH such as the high heating ratios which are reached, reduced processing time, energetic efficiency, formation of nano-sized grains in many cases, and so on. [8]

Thus, in this work we synthesized samples of the BSCCO system, focused on the 2212 phase. [9] We used the polymeric precursor method developed by Pechini, MP, associated to MAH. In one hand we worked with a chemical method which is a good synthesis route to produce homogeneous materials [1,3] and on the other hand we added a step on the synthesis process which consisted in the use of MAH.

## 2. Materials and methods

We focused on the synthesis of Bi-2212 due to its chemical stability and absence of toxic elements. We had already worked with different kind of chemical reagents, i.e., we used carbonates and nitrates. Thus, the samples were synthesized following PM and an association of PM with MAH. It was used a molar ratio of 3/1 for citric acid/metal and a mass ratio of 40/60 for citric acid/ethilenoglicol. [10].

A reference sample, REF, was synthesized by PM where we used carbonates as $(BiO_2)CO_3$, $SrCO_3$, $CaCO_3$ and $CuCO_3(OH)_2$ [11]. Another sample was also prepared using carbonates however, after the formation of the polymer, the precursor solution was submitted to MAH. The equipment used in this stage was a commercial microwave oven adapted with a Teflon® autoclave in which was inserted a collector cup. Such system is constituted by a thermocouple, a silicone sealing gasket, fixing screws, pressure gauge, safety valve and temperature controller. The synthesis was carried out using a heating rate of 2°C/min and maintained at 140°C for 60 minutes. During the heating process, the registered pressure was 3 bar. After that, the solution was kept in rest for during 24 hours and all the process was repeated once more. Thus, a supernatant material and a dark brown and viscous precipitate were originated. The precipitate was isolated and dried at 70°C during five days in an oven. The resulting sample was labeled as CMAH. A third sample was produced using nitrates as $Bi_5H_9N_4O_{22}$, $Sr(NO_3)_2$, $Ca(NO_3)_2$ and $Cu(NO_3)_2 \cdot 3H_2O$ for which was followed the same procedure as described above to obtain CMAH. This sample was labeled NMAH.

All powders were calcinated at 200°C/1h and 400°C/2h. After the calcinations, the powder was heat treated at 850°C [12] in two steps, one of them for 2h and the other one by 6h. The heating rate used was 2°C/min. The materials were characterized by x-ray diffraction, XRD, (Shimatzu model XRD-6000) and scanning electron microscopy, SEM, (ZEISS model EVO LS15). The XRD measurements were done in the 2θ ranged from 4° to 60° with 1°/min, steps of 0.02° and λ=1.542 Å with CuKα radiation.

The powders were pelletized and sintered at 845°C for 24h in air. After that, the samples were characterized by XRD, SEM and electric and magnetic measurements. The four-dc probe method was used to do the electric measurements which were carried out in a home-made system consisted of a current source Keithley 228A; a multimeter Keithley 2000, a nanovoltmeter Keithley 2182 and a Dewar with liquid nitrogen. The magnetic measurements were carried out in Quantum Design PPMS model 6000.

## 3. Results and discussion

### 3.1. XRD and SEM characterizations

The XRD of REF sample indicates the presence of the Bi-2212 phase, as shown in Figure 1(a). For the CMAH sample, the XRD data show two phases, Bi-2212 as major phase and Bi-2201 as second phase (see Figure 1(b)). The sample NMAH has the Bi-2212 phase with traces of Bi-2201 phase and other spurious compositions such as $SrCuO_2$ and $Bi_2Sr_3O_6$. Those analysis were made using JCPDS cards or equivalent.

Figure 2 shows SEM images of the as obtained powders and pellets using REF, CMAH and NMAH processes. In all cases were observed plate-like grains that are characteristic of the BSCCO superconducting system [13]. Figure 2 shows SEM micrographs of the pelletized samples. As can observed, in this figure, the REF sample presents plate-like grains with dimensions higher than 2x10 μm² and several small rounded grains with minimum dimensions around 1x2 μm²; in CMAH sample micrograph could be identified few grains with minor sizes of the order of 1x1 μm² even with thickness around 250 nm and the NMAH present plates higher than 2x4 μm² and few long needles with dimensions around 1.3x16 μm².

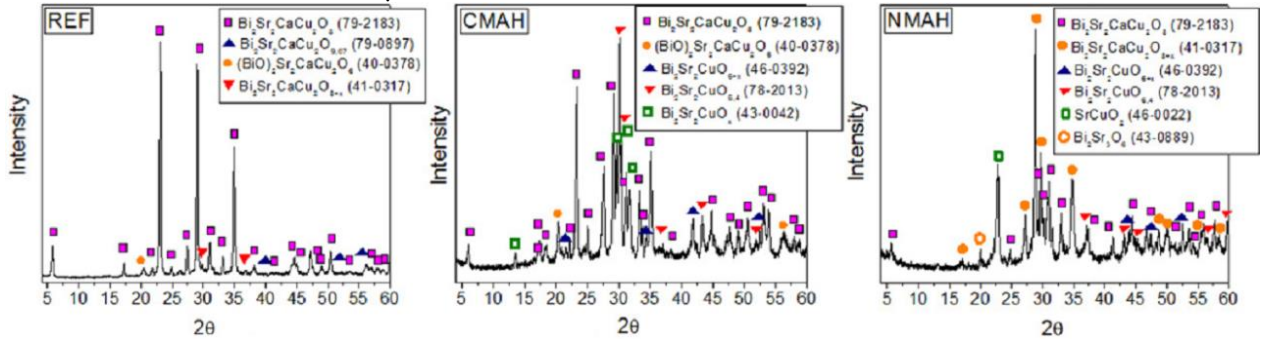

**Figure 1.** X-ray diffractogram of the samples REF, CMAH and NMAH.

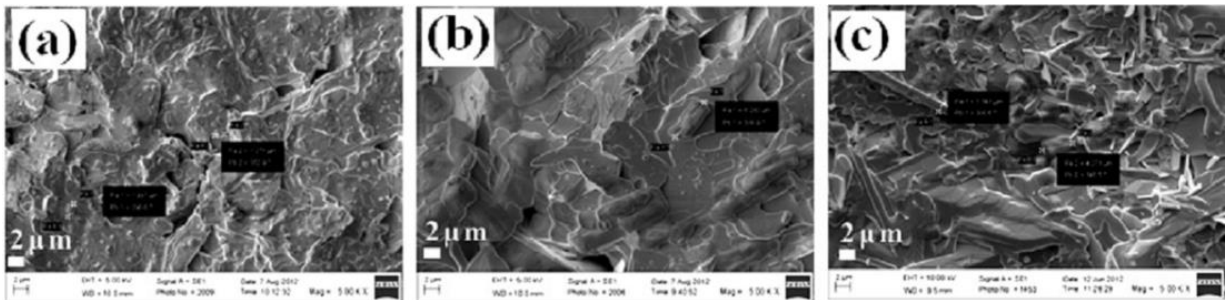

**Figure 2.** SEM micrographs of the pelletized samples (a) REF, (b) CMAH and (c) NMAH.

### 3.2. Electric and magnetic characterizations

Figure 3 shows the R(T) curves of the samples. In the worked range of temperatures can note some transitions in REF at 86K, 81K and 78K. The CMAH presents a transition at 81K and the NMAH sample also presents several transitions at 92K, 90K, 85K and 80K. For all samples can note the $T_c$ signature of the Bi-2212, as observed in the XRD analysis. The several transitions presented by all samples could be due to the presence of other superconducting phases with different $T_c$'s.

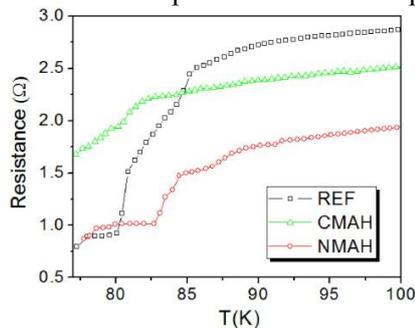

**Figure 3.** (Color online) Resistance as a function of the temperature curves of the samples REF, CMAH and NMAH. The several transitions presented by each sample is an indicative of the presence of different superconducting phases.

Figure 4 shows the M(T) curves of the samples. In all curves the $T_c$ is around 80K, characteristic of the Bi-2212 phase. It should be noted that the zero field cooled, ZFC, curve of REF has a greater diamagnetic response, in modulus, than for CMAH. Such behaviour is an indicative that the intergranular current flows through REF easier than in the CMAH. We can also note that the NMAH sample has a response quite different from that presented by REF, probably associated to its large grain-size distribution observed in SEM micrographs. However, the magnitude of its magnetization signal is comparable with that of the REF sample.

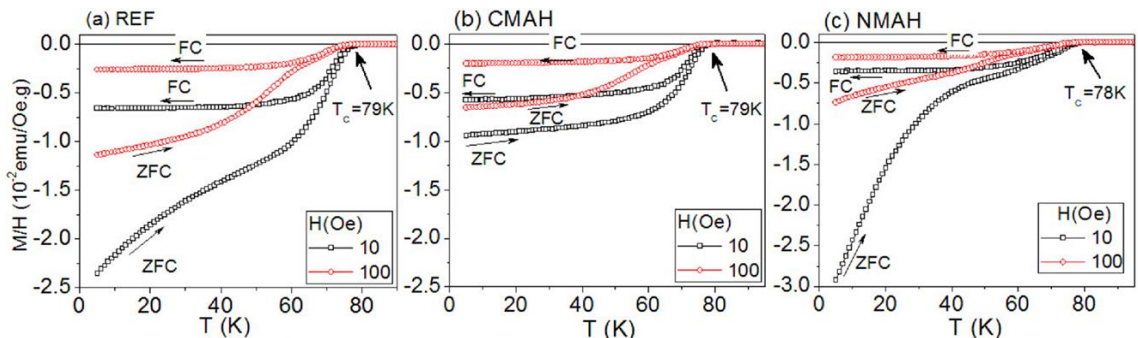
**Figure 4.** Magnetization versus temperature curves of the studied samples.

As conclusions we can note that both methods used in the synthesis of the samples produced the Bi-2212 phase with segregate phases. SEM micrographs show the presence of at least two type of grains, i.e., a plate-like grains which are characteristic of BSCCO superconducting system and rounded grains. In the electric characterizations, the several transitions presented by some samples could be related to other superconducting phases. The differences between electric and magnetic responses might be related to the fact that while the surface response is predominant in electric characterizations the magnetic ones measure an average response of the bulk sample. The measurements of the magnetization versus temperature followed the ZFC and FC procedures indicate that REF and NMAH samples could have grains with similar response under an external field. However the different behaviour of the M(T) curves could be associated with the grain-size distribution which is non-homogeneous in NMAH sample. Thus, it can be concluded with those analyses that the sample produced only by Pechini´s method present a better superconducting response than the samples produced by MAH. However, between MAH samples, the sample produced by nitrates, NMAH, presented a behaviour closer to that presented by REF sample. It is worth to emphasize that this is an initial study of the production of oxide superconductors by the microwave-assisted hydrothermal method.

We thank the Brazilian Agencies CAPES, CNPq, Fundunesp/PROPe and the São Paulo Research Foundation (FAPESP), grant 2013/11114-7, for financial support.

**References**
[1] Rao, C N R, Nagarajan R. and Vijayaraghaven R 1994 *Superc. Sci. Techn.* **6** 1
[2] Deimling C V, Motta M, Lisboa-Filho P N, Ortiz W A 2008, *J. Mag. Mag. Mat.* **320** e507
[3] Zhang Y, Yang H, Li M, Sun B and Qi Y 2010 *Cryst. Eng. Comm.* **12** 3046
[4] Keyson D, Longo E, Vasconcelos J S, Varela J A, Éber S, Dermaderosian A 2006 *Cerâmica* **52** 321
[5] Volanti D P, Cavalcante L S, Keyson D, Lima R C, Moura A P, Moreira M L, Macario L R, Godinho M, 2007 *Metalurgia & Materiais* **63** 351
[6] Baghurst D R, Chippindale A M and Mingos D M P 1988 *Nature* **332** 311
[7] Souza A E, Silva R A, Santos G T A, Moreira M L, Volanti D P, Teixeira S R, Longo E 2010 *Chemical Physics Letters* **488** 54
[8] Zhu X H and Hang Q M 2013 *Micron* **44** 21
[9] Maeda H, Tanaka Y, Fukutumi M and Asano T 1988 *J. Appl. Phys* **27** L209
[10] Chu C T and Dunn B 1987 *J. Am. Ceram. Soc.* **70** C-375
[11] Kakihana M 1996 *Sol-Gel Sci. Tech.* **6** 7
[12] Peng Z S, Hua Z Q, Li Y N, Di J, Ma J, Chu Y M, Zhen W N, Yang Y L, Wang H J and Zhao Z X 1998 *Journal of Superconductivity* **11** 6
[13] Majewski P 2000 *Journal of Materials Research* **15** 4